\documentclass[twocolumn,showpacs,preprintnumbers,amsmath,amssymb]{revtex4}
\usepackage{graphicx}
\usepackage{dcolumn}
\usepackage{bm}

\newcommand{\beq}{\begin{equation}}
\newcommand{\eeq}{\end{equation}}
\newcommand{\bea}{\begin{eqnarray}}
\newcommand{\eea}{\end{eqnarray}}

\begin{document}

\title{Pion and kaon parton distribution functions in a meson cloud model}

\author{C. Avila}\email{cavila@uniandes.edu.co}
\affiliation{Depto. de F\'{\i}sica, Universidad de los Andes,\\
AA 4976, Santaf\'e de Bogot\'a, Colombia}

\author{J. Magnin}\email{jmagnin@cbpf.br}
\affiliation{Centro Brasileiro de Pesquisas F\'{\i}sicas,\\
Rua Dr. Xavier Sigaud 150 - Urca - 22290-180\\
Rio de Janeiro - Brazil}

\author{J.C. Sanabria}\email{jsanabri@uniandes.edu.co}
\affiliation{Depto. de F\'{\i}sica, Universidad de los Andes,\\
AA 4976, Santaf\'e de Bogot\'a, Colombia}

\date{\today}

\begin{abstract}
In a model of hadrons in which the hadron wave function at low $Q^2$ is 
represented by a hadron-like Fock state expansion, we calculate the $\pi$ and $K$ 
parton distribution functions by comparing to experimental data. We show that the model 
gives an accurate description of available data and, in addition, gives a clear 
prediction of the initial flavors to be considered at the low $Q^2$ scale 
where perturbative evolution starts
\end{abstract}


\maketitle

The parton content of mesons is not well known due to the scarce experimental 
information as compared to the rich and accurate data existing for the proton. The 
meson structure is measured in dilepton Drell-Yan (D-Y) production in meson-nucleon interactions. 
So far, a few experiments\cite{pions,e615} have measured the valence quark distribution in 
charged pions, while for kaons, the only experimental information today available is on the 
$\bar{u}_k$ distribution in the $K^-$, which has been obtained by the NA3 
Collaboration~\cite{na3} in the 80's. 

In D-Y dilepton production in the reaction $H^A + H^B \rightarrow l^+~l^-$, the biggest 
contribution comes from the subprocesses $\bar{q}+q \rightarrow l^+~l^-$, where 
$\bar{q},~q$ are quarks in the initial hadrons $H^A,~H^B$. Then the partonic structure 
of one of the initial hadrons can be extracted if the structure of the other one is known. 
Thus, the structure of charged pions can be measured in $\pi^\pm N$ interactions. 
Furthermore, some authors have claimed that adequate linear combinations of $\pi^\pm p$ and 
$\pi^\pm n$ D-Y cross sections can allow the measurement of the valence quark distributions 
in charged pions independently of their sea quark and gluon distributions~\cite{some}.
In a similar way, the $\bar u$ valence quark distribution in the $K^-$ can be measured. However, 
the valence $s$ quark distribution cannot be easily determined in kaons since it has to anihilate 
with a $\bar s$ sea quark in the nucleon to produce the $l^+l^-$ pair, giving a negligible 
contribution to the D-Y cross section in the experimentally accessible kinematic region.

Then, to know the kaon structure, which is desirable since, for instance, kaons are frequently 
used as beam particles in experiments, one has to rely on theoretical models to 
complement the experimental information. On this respect, several models have been proposed. 
Among them we can mention the model of Ref.~\cite{hwa}, where the kaon structure is derived in a 
constituent quark model; the model of Ref.~\cite{grs}, where some quark and gluon sea is 
considered in addition to the constituent quarks at the scale where evolution starts; and most 
recently, the model presented in Ref.~\cite{ams}, where an expansion of the pion and kaon 
wave-functions in terms of hadron-like Fock states is done at the input $Q^2$ scale. 
One interesting feature of the last one is that the model predicts the parton 
structure of the pion and kaon, at the low $Q^2$ scale where evolution starts, up to a few 
parameters which must be fixed from experimental data. This is a remarkable property since 
leaves the model free of the arbitrariness of the other two, where by no means both the form and 
initial parton distribution functions (PDF) are justified.

In what follows, after a short revision of the model of Ref.~\cite{ams}, we shall use it 
to extract information on the pion and kaon parton content from simultaneous fits to the 
available experimental data. 

Following Ref.~\cite{ams}, the $\pi^-$ and $K^-$ wave functions can be written as
\bea
\label{eq1a}
\left|\pi^-\right> &=& a_0^{\pi}\left|\pi^-\right> + a_1^{\pi}\left|\pi^-g\right> 
+ a_2^{\pi}\left|K^0 K^-\right> + ... \\
\left|K^-\right> &=& a_0^K\left|K^-\right> + a_1^K\left|K^-g\right> 
+ a_2^K\left|{\overline K}\,^0 \pi^-\right> + ...\, ,
\label{eq1b}
\eea
at some low $Q_v^2$ scale, where we have neglected higher order contributions involving 
heavier mesons and fluctuations to Fock states containing more than two mesons. These 
fluctuations should be far off-shell and they can be safely ignored at this
point. 
The first terms in Eqs.~(\ref{eq1a})-(\ref{eq1b}) are the bare meson 
states, which are formed only by valons~\cite{hwa}. The following terms are 
fluctuations whose origin can be traced back to the processes $v\rightarrow v+g$ and 
$g\rightarrow q+\bar q$ followed by the recombination of the perturbative $q~\bar q$ pair with 
the valons, $v$, to form the hadron-like structure. As discussed in Ref.~\cite{cm2}, we assume 
that interactions between a quark or antiquark and a valon of the same flavor do not form a 
neutral, unflavored, virtual meson structure but annihilates nonperturbatively to a gluon. 
Additionally, in-meson hadrons are assumed to be formed only by valons.

Thus, these fluctuations are an effective representation of the nonperturbative -{\em intrinsic}- 
sea of quarks and gluons which should provide the necessary binding among constituent quarks to 
form hadrons~\cite{cm2,bps}. DGLAP evolution to higher $Q^2$ generates the perturbative 
-{\em extrinsic}- sea of $q\bar{q}$ pairs and gluons.

It is worth noting that, on very general basis, individual hadrons in the 
$\left|MM'\right>$ Fock states are colored. The same is also true for the $\left|M~g\right>$ 
fluctuations in the second term of Eqs.~(\ref{eq1a})-(\ref{eq1b}) as far as the gluon is 
in a color octet state. However, the fluctuation itself is colorless. Notice that, 
if the two components of a generic fluctuation are in the $\bm 8$ representation of 
color SU(3), and since  $\bm 8 \otimes 8 = 1 \oplus 8 \oplus 8 \oplus ...$, 
then there is a singlet (colorless) representation where the fluctuation can be accommodated.

Coefficients $a_i^M$; $M=\pi,K$; $i=1,2,3,...$; in Eqs.~(\ref{eq1a})-(\ref{eq1b})  
are constrained by probability conservation: $\sum_i{|a_i^M|^2} = 1$.

Since valon distributions in pions and kaons are related by
\bea
v_\pi(x) &\equiv & v_{\bar{u}/\pi^-}(x) = v_{d/\pi^-}(x) \nonumber \\
&=& v_{u/\pi^+}(x) = v_{\bar{d}/\pi^+}(x) \label{eq4a} \\
v_k(x) &\equiv & v_{\bar{u}/K^-}(x) = v_{d/K^\circ}(x) \nonumber \\
&=& v_{\bar{d}/{\overline K}^\circ}(x) = v_{u/K^+}(x) \label{eq4b} \\
v_{s/k} &\equiv & v_{s/K^-}(x) = v_{\bar{s}/K^+}(x) \nonumber \\
&=& v_{s/\overline{K}^\circ}(x) = v_{\bar{s}/K^\circ}(x) \label{eq4c}\; ,
\eea
due to isospin invariance, then at the $Q_v^2$ scale, parton distribution functions 
can be written as
\bea
\bar{u}_{\pi}(x) &=& d_{\pi}(x) = |a_0^\pi|^2 v_{\pi}(x) + 
|a_1^\pi|^2 P_{\pi g} \otimes  v_{\pi} \nonumber \\
& & +~|a_2^\pi|^2 P_{KK} \otimes v_{k} \label{eq2a} \\
s_\pi(x) &=& \bar{s}_\pi(x)= |a_2^\pi|^2 P_{KK} \otimes v_{s/k} \label{eq2b} \\
g_\pi(x) &=& |a_1^\pi|^2 P_{g\pi}(x)
\label{eq2c}
\eea
for pions, and
\bea
\bar{u}_K(x) &=& |a_0^K|^2 v_{k}(x) + 
|a_1^K|^2 P_{Kg} \otimes  v_{k} \nonumber \\
& & +~|a_2^K|^2 P_{\pi K} \otimes v_{\pi} \label{eq3a} \\
s_K(x) &=& |a_0^K|^2 v_{s/k}(x) + |a_1^K|^2 P_{Kg} \otimes  v_{s/k} \nonumber \\
& & +~|a_2^K|^2 P_{K\pi} \otimes v_{s/k} \label{eq3b} \\
d_K(x) &=& |a_2^K|^2 P_{\pi K} \otimes v_{\pi} \label{eq3c} \\
\bar{d}_K(x) &=& |a_2^K|^2 P_{K\pi} \otimes v_{k} \label{eq3d} \\
g_K(x) &=& |a_1^K|^2 P_{gK}(x)
\label{eq3e}
\eea
for kaons, where
\beq
P_{MM'} \otimes v_{q/M} \equiv \int_x^1{\frac{dy}{y}P_{MM'}(y)
v_{q/M}\left(\frac{x}{y}\right)}
\label{eq5}
\eeq
is the probability density of the nonperturbative contribution to the parton distribution 
coming from the $\left|MM'\right>$ fluctuation~\cite{cm2,mc}. 

The meson probability density $P_{MM'}(x)$ in the $\left|MM'\right>$ fluctuation has been 
calculated in Refs.~\cite{cm2,mc}. It is given by
\beq
P_{MM'}(x) = \int_0^1{\frac{dy}{y}\int_0^1{\frac{dz}{z} F(y,z)R(x,y,z)}}
\label{eq6}
\eeq
with
\bea
F(y,z) &=& \beta yv_q(y)zq'(z)(1-y-z)^a \label{eq7a} \\
R(x,y,z) &=& \alpha \frac{yz}{x^2}\delta(1-\frac{y+z}{x}) \; .
\label{eq7b}
\eea
In Eqs.~(\ref{eq7a})-(\ref{eq7b}), $v_q$ and $q'$ are the valon and the quark or antiquark 
distributions which 
will form the meson $M$ in the $\left|MM'\right>$ fluctuation. The $q'$ distribution is 
generated through the gluon emission from a valon followed by the $q'~\bar{q}\,'$ pair 
creation. Thus its probability density is~\cite{cm}
\bea
q'(x) &=& \bar{q}'(x) = N \frac{\alpha_{st}^2(Q^2_v)}{(2\pi)^2}
\int_x^1 {\frac{dy}{y} P_{qg}\left(\frac{x}{y}\right)} \times  \nonumber \\
& &\int_y^1{\frac{dz}{z} P_{gq}\left(\frac{y}{z}\right) v_q(z)} \; ,
\label{eq9}
\eea  
where $P_{qg}(z)$ and $P_{gq}(z)$ are the 
Altarelli-Parisi splitting functions~\cite{a-p},
\bea
P_{gq} (z) &=& \frac{4}{3} \frac{1+(1-z)^2}{z},\label{eq10a} \\ 
P_{qg} (z) &=& \frac{1}{2} \left( z^2 + (1-z)^2 \right).
\label{eq10b}
\eea

The only scale dependence appearing in Eq.~(\ref{eq9}) arises 
through $\alpha_{st}(Q^2)$. Since the valon scale is tipically of 
the order of $Q_v^2\sim0.64$ GeV$^2$~\cite{hwa}, then the $q'~\bar{q}\,'$ pair creation can 
be safely evaluated perturbatively because $\alpha_{st}^2/(2\pi)^2$ is still sufficiently 
small. The normalization constants $\alpha$, $\beta$ and $N$ in Eqs.~(\ref{eq7a})-(\ref{eq7b}) 
and (\ref{eq9}) contribute to the global normalization coefficient of the corresponding Fock 
state fluctuation in the expansions of Eqs.~(\ref{eq1a})-(\ref{eq1b}).

Momentum conservation also requires 
\beq
P_{MM'}(x) = P_{M'M}(1-x) \; ,
\label{eq11}
\eeq
a condition which relates the in-meson $M$ and $M'$ probability densities.
Additionally, hadrons in the $\left|MM'\right>$ fluctuation must be correlated in
velocity in order to form a bound state. This imply that
\beq
\frac{\left<xP_{MM'}(x)\right>}{m_M} = \frac{\left<xP_{M'M}(x)\right>}{m'_M} \; ,
\label{eq12}
\eeq
fixing the exponent $a$ in Eq.~(\ref{eq7a}). Notice also that $P_{gM}$ 
is calculated from the ``recombination'' of an antiquark with a valon of the same 
flavor~\cite{cm2}. Then, formally, the $P_{gM}$ corresponds to the would be 
$\pi^0$ distribution in a hypothetical $\left|M\pi^0\right>$ fluctuation.

With the $\pi$ and $K$ PDF given in Eqs.~(\ref{eq2a})-(\ref{eq2c}) and 
Eqs.~(\ref{eq3a})-(\ref{eq3e}) we proceed to fit simultaneously the experimental data 
on the $\bar{u}_\pi$ and $\bar{u}_K/\bar{u}_\pi$ by the E615~\cite{e615} and NA3~\cite{na3} 
Collaborations respectively. To that end we parametrize~\cite{hwa}
\bea
v_\pi(x)   &=& \frac{1}{\beta(a_\pi+1,b_\pi+1)}x^{a_\pi}(1-x)^{b_\pi}\; , \label{eq13a}\\
v_k(x)     &=& \frac{1}{\beta(a_k+1,b_k+1)}x^{a_k}(1-x)^{b_k}\; , \label{eq13b}\\
v_{s/k}(x) &=& \frac{1}{\beta(a_k+1,b_k+1)}x^{b_k}(1-x)^{a_k}\; ; \label{eq13c}
\eea
where the last two are related one another by momentum conservation; and construct a $\chi^2$ 
function as
\bea
\chi^2 &=& \chi^2_{E615}+\chi^2_{NA3} \nonumber \\
&=&
\sum{\frac{\left[y_i-x_iu_\pi(x_i,Q^2_{E615})\right]^2}{\sigma_{y_i}^2}} \nonumber \\
&+& \sum{\frac{\left[z_i-\frac{u_k}{u_\pi}(x_i,Q^2_{NA3})\right]^2}{\sigma_{z_i}^2}}\;,
\label{chi2}
\eea
to be minimized. $\beta(a,b)$ in Eqs.~(\ref{eq13a})-(\ref{eq13c}) is the Beta
function, which gives the correct normalization.

We have a total of 8 free parameters to be fixed by the fit, namely, $\left|a_0^{\pi,K}\right|$, 
$\left|a_1^{\pi,K}\right|$ and the four exponents in the valon distributions of 
Eqs.~(\ref{eq13a})-(\ref{eq13c}). The fit procedure started from an arbitrary set
of parameters, for which the low $Q^2$ $\pi$ and $K$ PDF were calculated and
evolved to the $Q^2$ scale of the E615 and NA3 experiments respectively. Then
the $\chi^2$ was calculated. This procedure was repeated until a minimun
of $\chi^2$ was found.

\begin{figure}[ht]
\includegraphics[width=8.5cm]{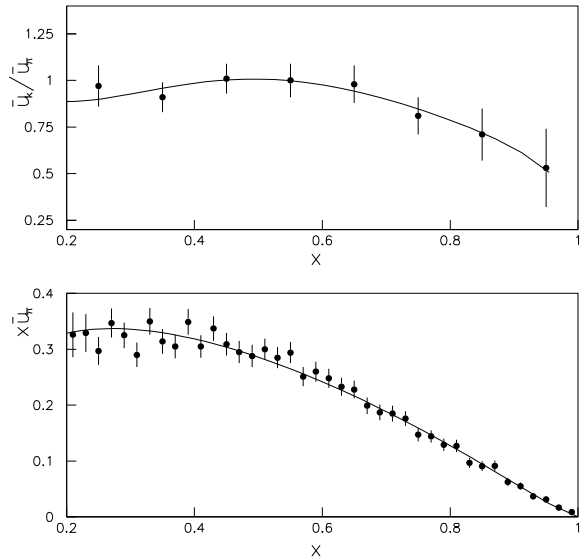}
\caption{\label{fig1} Fit to experimental data from the E615~\cite{e615} and NA3~\cite{na3} 
experiments. Upper: the $\bar{u}_k/\bar{u}_\pi$ ratio as a function of $x$ as measured by NA3. 
Lower: the $\bar{u}_\pi$ distribution as a function of $x$ as measured by the E615 Collaboration.}
\end{figure}

Results of the fit are shown in Table~\ref{table1} and displayed in 
Fig.~\ref{fig1}. As evidenced by the figure, the quality of the fit is quite good.

\begin{table}
\caption{\label{table1}Coefficients of the pion and kaon PDF obtained from simultaneous fits 
to experimental data from the E615~\cite{e615} and NA3~\cite{na3} Experiments. The 
$\chi^2/d.o.f.$ of the fit is $0.882$. Note that coefficients $\left|a_2^\pi\right|^2$ and 
$\left|a_2^K\right|^2$ are not independent but fixed by probability conservation. }
\begin{ruledtabular}
\begin{tabular}{lcr}
$\left|a_0^\pi\right|^2$ & 0.689 & $\pm$ 0.041\\
$\left|a_1^\pi\right|^2$ & 0.310 & $\pm$ 0.130\\
$\left|a_2^\pi\right|^2$ & 0.001 & $\pm$ 0.136\\
\hline
$\left|a_0^K\right|^2$   & 0.411 & $\pm$ 0.067\\
$\left|a_1^K\right|^2$   & 0.229 & $\pm$ 0.097\\
$\left|a_2^K\right|^2$   & 0.360 & $\pm$ 0.118\\
\hline
$a_\pi$                  & 0.044 & $\pm$ 0.036\\
$b_\pi$                  & 0.372 & $\pm$ 0.025\\
$a_k$                    & 0.917 & $\pm$ 0.898\\
$b_k$                    & 0.743 & $\pm$ 0.240
\end{tabular}
\end{ruledtabular}
\end{table}

\begin{figure}[ht]
\includegraphics[width=8.5cm]{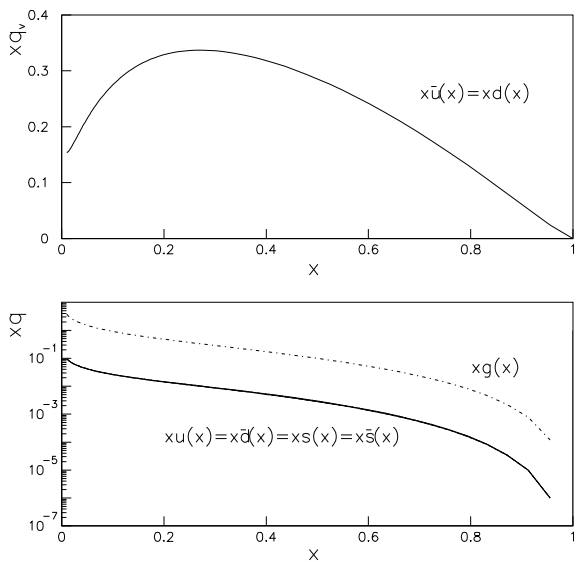}
\caption{\label{fig2} Parton distribution functions in charged pions at $30$ GeV$^2$.
Upper: valence quark distributions ($xd(x)=x\bar{u}(x)=x\bar{u}_v(x)+
x\bar{u}_s(x)$). Lower: sea quark $u_\pi=\bar{d}_\pi=s_\pi=\bar{s}_\pi$ (full line), 
and gluons (dot line).}
\end{figure}
\begin{figure}[ht]
\includegraphics[width=8.5cm]{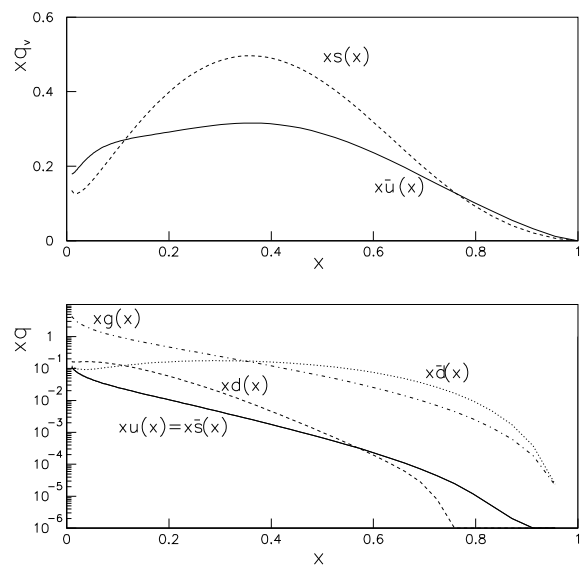}
\caption{\label{fig3} Parton distribution functions in kaons at $30$ GeV$^2$. 
Upper: valence quark distributions $\bar{u}_k=\bar{u}_v+\bar{u}_s$ (full line)
and $s_k=s_v+s_s$ (dashed line). 
Lower: sea quark and gluon distributions $u_k=\bar{s}_k$ (full line), $d_k$ (dashed line), 
$\bar{d}_k$ (dot dashed line) and gluons (dot line).}
\end{figure}

Once the parameters in the pion and kaon wave-functions are known, we build the full set 
of PDF. They are shown for both, the $\pi^-$ and the $K^-$, in Figs.~\ref{fig2} and 
\ref{fig3} respectively at $Q^2 = 30$ GeV$^2$. To obtain the pion and kaon PDF we have used 
the central values of parameters in Table~\ref{table1}. Note that, since $\left|a_2^\pi\right|^2
\sim 0$, then the quark sea of the pion is nearly SU(3)-flavor 
symmetric~\cite{foot1}. However, in the quark sea of 
the kaon the SU(3) symmetry is largely broken due to the big contribution of fluctuations 
to the intrinsic sea. Another interesting feature of the kaon structure is the
$d-\bar d$ asymmetry of the $K^-$ sea. This is due to the fact that $d$ is a
valon in the $\pi$ while $\bar d$ is a valon in the $K$ in the
$\left|K\pi\right>$ fluctuation. Isospin symmetry tell us that $(d-\bar
d)^{K^-}(x)=(\bar{d}-d)^{K^+}(x)=(u-\bar
u)^{\overline{K}^0}(x)=(\bar{u}-u)^{K^0}(x)$. The $u$ and $\bar s$ sea are, however,
  equal in the $K^-$. Once again, via the isospin symmetry, we get $\bar{u}=s$
  in the $K^+$ sea, $d=\bar d$ in the $\overline{K}^0$ sea and $\bar{d}=s$ in
  the $K^0$ sea~\cite{foot1}.

The PDF of $\pi^+$, $K^+$, $K^0$ and $\bar{K}^0$ can be obtained just by 
using isospin symmetry. The $\pi^0$ structure, however, deserves separate considerations. 
In fact, the $\pi^0$ wave-function at the low $Q_v^2$ scale can be writen as~\cite{cm2} 
\bea
\left|\pi^0\right> &=& b_0\left|\pi^0\right> + b_1 \left|\pi^0\,g\right> 
+ b_2 \left|\pi^-\,\pi^+\right> \nonumber \\
& &+ \frac{b_3}{\sqrt{2}} \left[ \left|K^-\,K^+\right> 
- \left|K^0\,\bar{K}^0\right> \right] + ... \; .
\label{eq15a}
\eea
Thus, contrarily to charged pions, where the intrinsic sea is formed only by 
gluons and eventually strange quarks, in the $\pi^0$ the intrinsic sea is formed also 
by $u\bar u$ and $d\bar d$ quarks due to the $\left|\pi^+\pi^-\right>$ fluctuation.
It should be noted also that, as the $\left|\pi^0\,g\right>$ and the 
$\left|\pi^+\pi^-\right>$ fluctuations have the same origin, namely the splitting of a 
gluon to a $u\bar u$ or $d\bar d$ pair, then it can be assumed that $|b_1|^2 \sim |b_2|^2$, 
thus possibly reducing the intrinsic gluon sea due to probability conservation. 
This indicates a remarkable difference between the structure of charged and neutral 
pions within the model.

We want to stress that we have determine the complete structure of charged pions and kaons 
from a minimal set of measurements of the $\pi^-$ and $K^-$ PDF just by extracting the 
three valon distributions $v_\pi$, $v_k$ and $v_{s/k}$ and the parameters 
$\left|a_0^{\pi,K}\right|^2$ and $\left|a_1^{\pi,K}\right|^2$ from experimental data.
In this sense the model has an interesting predictive power. Note that, as a matter of fact, the 
experimental information one can get on the kaon structure is only on the light valence quark 
distribution. The measurement of the strange and even sea quark distributions in kaons 
is not possible due to practical reasons. Actually, strange and sea quarks only contribute 
to the total Drell-Yan dilepton cross section through valence-sea and sea-sea $q\bar{q}$ 
anihilation. Then their contributions are small and cannot be easily separated. 
A similar difficulty arises to measure the sea parton distributions in pions, where, once 
again, valence-sea and sea-sea contributions to the Drell-Yan cross section are small than 
the valence-valence ones. 

Moreover, the model predicts the structure of pions and kaons at the low $Q_v^2$ 
scale, where perturbative QCD evolution starts. This gives a plausible solution to the long 
standing problem of the origin of the valence-like sea quark and gluon
distributions nedeed, at the low $Q^2$ input scale for evolution, to describe
experimental data on hadron structure.

\begin{acknowledgments}
J.M. is grateful for the warm hospitality at the Physics Department, Los Andes University, where 
part of this work was done.
\end{acknowledgments}


\begin{thebibliography}{99}
\bibitem{foot1} This statement is valid only if the probability
  of the gluon splitting to a $q~\bar{q}$ pair is the same irrespective of the
  quark flavor.
\end{thebibliography}
\end{document}